\documentclass[aip, jcp, twocolumn, floatfix, groupaddress, reprint, 10pt]{revtex4-1}  
\usepackage{bm}
\usepackage{graphicx}
\usepackage[usenames]{color}
\usepackage{microtype}
\usepackage{amsmath}
\usepackage{amssymb}
\usepackage{xspace}
\usepackage{footnote}
\usepackage{relsize}
\usepackage{hhline}

\usepackage{physics}
\usepackage{hyperref}
\usepackage[normalem]{ulem}

\newcommand{\br}{\boldsymbol{r}}

\newcommand{\bk}{\boldsymbol{k}}

\usepackage{algorithm}

\begin{document}

\title{Accelerating QM/MM simulations of electrochemical interfaces through machine learning of electronic charge densities
}

\author{Andrea Grisafi}
\email{andrea.grisafi@sorbonne-universite.fr}
\affiliation{Institut Sciences du Calcul et des Données, ISCD, Sorbonne Universit\'e, F-75005 Paris, France}

\author{Mathieu Salanne}
\email{mathieu.salanne@sorbonne-universite.fr}
\affiliation{Physicochimie des \'Electrolytes et Nanosyst\`emes Interfaciaux, Sorbonne Universit\'e, CNRS, F-75005 Paris, France
}\affiliation{{Institut Universitaire de France (IUF), F-75231 Paris, France}}

\begin{abstract}
A crucial aspect in the simulation of electrochemical interfaces consists in treating the distribution of electronic charge of electrode materials that are put in contact with an electrolyte solution. 
Recently, it has been shown how a machine-learning method that specifically targets the electronic charge density, also known as SALTED, can be used to predict the long-range response of metal electrodes in model electrochemical~cells. In this work, we provide a full integration of SALTED with MetalWalls, a program for performing classical simulations of electrochemical systems. We do so by deriving a spherical harmonics extension of the Ewald summation method, which allows us to efficiently compute the electric field originated by the predicted electrode charge distribution. We show how to use this method to drive the molecular dynamics of an aqueous electrolyte solution under the quantum electric field of a gold electrode, which is matched to the accuracy of  density-functional theory. Notably, we find that the resulting atomic forces present a small error of the order of 1~meV/{\AA}, demonstrating the great effectiveness of adopting an electron-density path in predicting the electrostatics of the system. Upon running the data-driven dynamics over about 3~ns, we observe qualitative differences in the interfacial  distribution of the electrolyte with respect to the results of a classical simulation. By~greatly accelerating quantum-mechanics/molecular-mechanics approaches applied to electrochemical systems, our method opens the door to nanoseconds timescales in the accurate atomistic description of the electrical double layer. 
\end{abstract}

\maketitle

\section{Introduction}

Understanding the interface between a metallic electrode and a liquid electrolyte has become an important objective in chemical physics.~\cite{Dawlaty2023} This interest is driven by the many underlying applications in electrochemistry, electrocatalysis, and energy storage, but it is also a challenging problem from a fundamental point of view. Most of the time, the conventional simulation methods cannot easily be adopted due to the highly heterogeneous nature of the system, the difficulties associated with the application of a voltage between two electrodes, and the large time and length scales required to simulate a representative portion of the electrochemical interface.~\cite{Schwards2020,gross2023}

In the absence of chemical reactions, intermolecular interactions in bulk electrolytes are often well handled using molecular mechanics (MM) approaches.~\cite{Goloviznina2019} The main difficulty is therefore to account for the electronic structure of the metal. For this purpose, performing quantum mechanics (QM) calculations, most notably at the density functional theory (DFT) level, appears as the most accurate available approach. Although it is possible to use DFT to calculate atomic forces within a molecular dynamics (MD) simulation, the large computational cost limits its application to relatively simple systems.~\cite{Li2019natmat,Khatib2021,Chen2022} Typically, having an electrolyte including a salt at finite concentration remains a hurdle for such simulations. A possible solution is to use a purely classical approach, which leads to the family of constant-potential MD simulations.~\cite{Jeanmairet2022} In this case, the polarization of the metal is treated using fluctuating atomic Gaussian charges, which depend on both the electrolyte positions and the applied potential. This approach was refined to effectively include some electronic structure effects,~\cite{Scalfi2020thomas-fermi,goloviznina2024b} but the electrode representation remains highly simplified when compared to an explicit QM-based method.

An alternative is to mix the QM/MM approaches by partitioning the system. In fact, by performing the QM calculation on the electrode subsystem only, it is possible to greatly decrease the computational burden. Such calculations were for example used to study supercapacitors,~\cite{elliott2023a} the electrowetting of graphene,~\cite{wei2024} or the electrocatalytic reactivity of MoS$_2$-based materials.~\cite{abidi2023a} There are however several limitations. Firstly, performing a QM calculation at every timestep of a MD simulation ($\approx$~1fs) is not very efficient, so that multi-stepping techniques have to be introduced by updating the QM system every few thousands steps. Secondly, while the energy of the system can be recovered by adequately summing up the energies of the sub-systems, the same is not always true for the forces. In particular, calculating the electrostatic forces acting on the MM subsystem due to the QM subsystem is not straightforward. The problem is generally tackled by introducing equilibrated partial charges on the electrode atoms,~\cite{Takahashi2022} which are computed following a quadratic functional form of the electrostatic energy. Beside the assumed quadratic approximation, this type of charge partitioning is not rigorously defined, so that some important physical effects associated with the local anisotropy of the charge distribution may be lost in the process.

In this work, we introduce a data-driven method that allows us to improve both the efficiency and the accuracy of QM/MM simulations of electrochemical interfaces. This method is grounded on recent developments in a machine-learning (ML) approach capable of predicting the electronic charge density of metal electrodes, including its non-local response to far-field perturbations.~\cite{grisafi2023prm} In particular, we implement an interface between the SALTED ML program~\cite{salted} and the MetalWalls MD code for electrochemical simulations.~\cite{coretti2022a} As a key aspect of this interface, we provide an efficient calculation of the electric field generated by the predicted electronic charge distribution, which we derive following a spherical harmonics generalization of the Ewald summation method. In turn, this allows us to compute accurate electrostatic forces acting on the electrolyte atoms, positioning our method in stark contrast with approaches that are based on partial charges. By shifting the largest share of the computational costs on the learning stage, simulation times of the order of nanoseconds can be achieved with a relatively small overhead compared to conventional full MM simulations. As an application of our method, we study the interface between a gold electrode and an aqueous solution of NaCl at 5~M concentration under applied electric potentials of 0 and $1$~V. The structure of the liquid phase as well as interfacial properties such as the capacitance are calculated and compared to previous approaches.

\section{General method\label{sec:method}}

We adopt a finite-field simulation setup, where a single electrode is put in contact with an electrolyte solution under three-dimensional periodic boundary conditions. This setup can be conveniently used to simulate a pair of equivalent electrodes that are kept at a constant potential difference.~\cite{Dufils2019} In particular, a uniform and constant electric field, $\varepsilon_z$, is applied along the direction normal to the electrode surface, such that $\Delta V = -\varepsilon_z\, L_z$, with $L_z$ the length of the simulation box, corresponds to a physical electrochemical cell potential. The representation of the electrode/electrolyte interface follows the approach already presented in Ref.~\citenum{grisafi2023prm}. Specifically, we  adopt a quantum-mechanics/molecular-mechanics (QM/MM) strategy, where an \textit{ab initio} representation of the electrode is used in combination with a classical representation of the electrolyte defined via fixed Gaussian charges. We aim to rely on SALTED predictions of the electrode charge density in order to compute accurate electrostatic forces that can be used to drive the dynamics of the system. 

\subsection{Representation of the electrode charge density}

Within SALTED, the electronic charge density of the electrode is represented as a linear spherical harmonics expansion that follows density-fitting (DF) approximations commonly adopted in electronic-structure methods:~\cite{Grisafi2023,lewis+21jctc}
\begin{equation}\label{eq:density}
    n_e(\br) = \sum_{inlm} c^{nlm}_i \sum_{\boldsymbol{u}} R_{nl}(\left|\br-\br_i-\boldsymbol{u}\right|)Y^{l}_m(\widehat{\br-\br_i-\boldsymbol{u}})\, .
\end{equation}
Here, the index $i$ identifies the electrode atoms, $nlm$ the basis function indexes, while $\boldsymbol{u}$ represent the three-dimensional translation vectors over the periodic images of the system. We assume that radial auxiliary functions are of Gaussian-type, i.e., $R_{nl} \propto r^l e^{-r^2/(2\sigma^2_n)}$, while $Y^{l}_m$ are defined as orthonormalized real spherical harmonics. 

Reference calculations of $n_e$ in the presence of the classical electrolyte field are performed using the CP2K simulation program.~\cite{cp2k2020} Specifically, quantum-mechanical calculations are performed at the Kohn-Sham DFT level using the PBE functional~\cite{perd+96prl} with DZVP-MOLOPT-SR basis sets~\cite{VandeVondele+07jcp} and GTH pseudo-potentials.~\cite{goedecker1996prb} To extract the expansion coefficients, $c^{nlm}_i$, we adopt a density-fitting approximation that makes use of a truncated Coulomb metric,~\cite{Guidon2009,Bussy2023} a choice that is especially suitable to  represent the electrostatic properties of the system.

The total electrode charge density is obtained by adding to $n_e$ the effective nuclear-charge density associated with the classical local part of the GTH pseudopotential. For each electrode atom $i$, this is simply defined by an isotropic Gaussian charge, i.e., 
\begin{equation}\label{eq:pseudo-charge}
    n_\text{ion}(r) = \frac{Z_\text{eff}}{\sqrt{8}\, \pi^{3/2}\, r^3_\text{loc}} \exp\left[-\frac{1}{2}\left(\frac{r}{r_\text{loc}}\right)^2\right]\, ,
\end{equation}
where $Z_\text{eff}$ is the effective nuclear charge of the electrode atoms and $r_\text{loc}$ is the local pseudopotential cutoff.

\subsection{SALTED electrode model}

In analogy with the prescription adopted in Ref.~\citenum{grisafi2023prm}, we will assume that the positions of the electrode atoms are kept fixed, so that the learning can be focused on the sole electrolyte-induced variations of the electrode charge density. Moreover, as the applied voltage $\Delta V$ is typically of the order of a few volts over a cell that covers several nanometers along $z$, we can further assume that the external electric field, defined as $\varepsilon_z \equiv -\Delta V/L_z$, introduces a small linear perturbation to $n_e$. Under these assumptions, we can conveniently define the SALTED prediction target as the following density difference,
\begin{equation}
    \Delta n_e(\br) = n_e(\br;\varepsilon_z) - n^0_e(\br;\varepsilon_z)\, , 
\end{equation}
where $n^0_e$ is the electron density of the isolated electrode under $\varepsilon_z$, used as a constant baseline for the regression procedure. Upon training, $n^0_e$ is eventually added back to the predicted $\Delta n_e$ in order to recover the total electron density $n_e$. Notably, the possibility of treating the external field as a small perturbation implies that $\Delta n_e$ can be considered independent from $\varepsilon_z$. From a ML point of view, this represents a great advantage, as it allows us to simulate a continuum spectrum of cell potentials without the need to retrain a new SALTED model whenever a different value of $\varepsilon_z$ is applied. 

Consistently with the density-fitting metric used to generate the reference quantum-mechanical data, SALTED models can be trained by minimizing a loss function that makes use of a (truncated) Coulomb metric.~\cite{briling2021} In compact notation, this reads as follows,
\begin{equation}\label{eq:loss}\begin{split}
\mathcal{L}(\boldsymbol{w}) &=  \sum_{I=1}^N \bra{n^\text{ML}_{e}(\boldsymbol{w})-n^\text{DF}_e} \frac{1}{\br-\br'}\ket{n^\text{ML}_{e}(\boldsymbol{w})-n^\text{DF}_e}_I 
\\&=\sum_{I=1}^N \left[\boldsymbol{c}^\text{ML}_I(\boldsymbol{w})-\boldsymbol{c}^\text{DF}_I\right]^T  \mathbf{J}_I  \left[\boldsymbol{c}^\text{ML}_I(\boldsymbol{w})-\boldsymbol{c}^\text{DF}_I\right]\, ,
\end{split}\end{equation}
where $N$ is the number of training configurations. In the second line of this equation, $\mathbf{J}$ is the matrix of 2-centers Coulomb integrals between the auxiliary functions, i.e., $\bra{i'n'l'm'}r^{-1}\ket{inlm}$, as computed by CP2K, $\boldsymbol{c}^\text{DF}$ and $\boldsymbol{c}^\text{ML}$ are the density-fitting and machine-learning vectors of density coefficients associated with the expansion of~$\Delta n_e$, respectively, and $\boldsymbol{w}$ is the vector of regression weights. In practice, a regularization term is also added to the loss function to avoid overfitting; we refer to Ref.~\citenum{Grisafi2023} for an in-depth discussion of the SALTED method.

Following Ref.~\citenum{grisafi2023prm}, the symmetry-adapted kernels used for the approximation of the $\boldsymbol{c}^\text{ML}$ density coefficients can be  defined to include long-range information. In particular, kernel functions between atomic environments are constructed from long-distance equivariant (LODE) structural descriptors,~\cite{gris-ceri19jcp,grisafi2021cs,Huguenin-Dumittan2023} as implemented in the ``rascaline'' atomistic representations package.~\cite{rascaline} This feature of the method is fundamental to capture the non-local variations of the electron density induced by electrolyte charges that are located arbitrarily far from the electrode surface.~\cite{grisafi2023prm} We~remark that, unlike charge equilibration schemes, this is made possible without relying on any self-consistent procedure. In fact, the use of long-range descriptors allows us to learn directly the self-consistent electronic polarization of the system, thus bypassing the need to incorporate self-consistency at the ML level. This would still hold true in the scenario where the electrolyte is also undergoing an electronic polarization. 

\section{Electric field calculation\label{sec:efield}}

From the predicted electrode charge distribution, a nontrivial problem is that of performing an efficient calculation of the electric field needed to drive the dynamics of the electrolyte. To achieve this goal, we provide in this section an original analytical derivation of the Ewald summation method, properly extended to account for the spherical harmonics expansion of the electron density.

Following the representation of Eq.~\eqref{eq:density}, we start by introducing a fictitious electron density, $n^\text{ewald}_e$, obtained by combining the physical density coefficients, $c^{nlm}_i$,  with a new set of smooth radial functions, $R_{nl}^\text{ewald}$. Crucially, these functions are defined from a Gaussian width $\sigma_\text{ewald}$ that is larger than those entering the definition of $R_{nl}$. 
Following the conventional Ewald summation method, we then separate from the electron density a charge-neutral (screened) contribution, defined as 
\begin{equation}
    n^\text{screen}_e(\br) \equiv n_e(\br)-n^\text{ewald}_e(\br)\, .
\end{equation}
By construction, the electric field generated by $n^\text{screen}_e$ is short-ranged, and can be efficiently computed in real space. Conversely, the remaining (unscreened) $n^\text{ewald}_e$ density will contribute to a long-range electric field that can be efficiently computed in reciprocal space. In what follows, we outline the derivation of both electronic and nuclear contributions to the electric field, as newly implemented in the MetalWalls simulation program. 

\subsection{Short-range electronic contribution}

Starting from $n^\text{screen}_e$, the short-range part of the Hartree electronic potential can be analytically computed in real space by relying on the Laplace expansion of the Coulomb operator. After some calculations, detailed in the Supplementary Material, we obtain
\begin{equation}
    \begin{split}\label{eq:ewald-pot-real}
&\phi^\text{SR}_\text{H}(\br) = \int_{\mathbf{R}^3} d\br'\, \frac{n^\text{screen}_e(\br')}{|\br-\br'|}  
\\&=4\pi \sum_{i=1}^{N_\text{at}}\sum_{nlm} \frac{c_{i}^{nlm}}{2l+1} \sum_{\boldsymbol{u}}Y_{lm}(\widehat{\br-\br_i-\boldsymbol{u}})\\&\times\left[\frac{I^<_{nl}(|\br-\br_i-\boldsymbol{u}|)}{|\br-\br_i-\boldsymbol{u}|^{l+1}} + |\br-\br_i-\boldsymbol{u}|^lI^>_{nl}(|\br-\br_i-\boldsymbol{u}|)\right]\, .
\end{split}
\end{equation}
Here, $I_{nl}$ indicate radial integrals performed within ($<$) and outside ($>$) a sphere of radius $r\equiv|\boldsymbol{r}-\boldsymbol{r}_i-\boldsymbol{u}|$:
\begin{equation}\label{eq:radial_integrals}
\begin{split}
    &I_{l n}^<(r) = \int_0^r ds\,s^{2+l} \left[R_{nl}(s) - R^\text{ewald}_{nl}(s)\right]\, , \\&
    I_{l n}^>(r) = \int_r^\infty ds\,s^{1-l} \left[R_{nl}(s) - R^\text{ewald}_{nl}(s)\right]\, .
    \end{split}
\end{equation}
While the individual terms of $I_{l n}^>$ are quickly vanishing for $r\to\infty$, the requirement that $I_{l n}^<$ vanishes sufficiently fast for increasing values of $r$ implies that the normalization of the Ewald radial functions must be chosen so that to satisfy the following equality:
\begin{equation}
     \int_0^\infty ds\,s^{2+l} R^\text{ewald}_{nl}(s) = \int_0^\infty ds\,s^{2+l} R_{nl}(s)\, .
\end{equation}
This choice guarantees that both types of radial integrals in Eq.~\eqref{eq:radial_integrals} tend to zero with superexponential rapidity, thus dominating over the algebraic terms of the order of $1/r^{l+1}$ and $r^l$ in Eq.~\eqref{eq:ewald-pot-real}. Crucially, this screening effect allows us to compute the short-range part of the Hartree potential by only collecting contributions of electrode atoms that fall within a finite cutoff distance. In particular, we select a cutoff of $r_\text{cut}^l=4\sqrt{2+l}\, \sigma_\text{Ewald}$, which is defined to account for the spatial range covered by the Ewald Gaussian-type functions.

To compute the electric field as $\boldsymbol{E}^\text{SR}_\text{H} = -\nabla \phi^\text{SR}_\text{H}$, it is convenient to adopt the substitution $\boldsymbol{r}-\boldsymbol{r}_i-\boldsymbol{u} \equiv \left(r,\theta,\phi \right)$ in Eq.~\eqref{eq:ewald-pot-real}, so that to make use of the gradient in spherical coordinates, i.e.,
\begin{equation*}
    \tilde{\nabla} \equiv \left(\frac{\partial}{\partial r}, \frac{1}{r}\frac{\partial}{\partial \theta}, \frac{1}{r\sin\theta}\frac{\partial}{\partial \phi}\right)\, .
\end{equation*}
Upon this change of variables, the contribution to the short-range part of the electric field brought by each electrode atom can be analytically computed. In compact notation, we have
\begin{equation}\begin{split}
     \boldsymbol{E}^\text{SR}_\text{H}(\br)  &= 
        4\pi\sum_{i=1}^{N_\text{at}} \sum_{nlm} \frac{c_i^{nlm}}{2l+1}\,   \sum_{\boldsymbol{u}}\mathbf{U}(\widehat{\br-\br_i-\boldsymbol{u}}) \\&
 \cdot  \tilde{\nabla} \left[ \left(\frac{1}{r^{l+1}}\,I_{l n}^<(r) + r^{l}\, I_{l n}^>(r)\right)Y^l_m(\theta,\phi)\right]\, ,
\end{split}
\end{equation}
where
\begin{equation}\label{eq:unimatrix}
 \mathbf{U} =    \begin{pmatrix}
 \sin\theta\cos\phi& \cos\theta\cos\phi & -\sin\phi\\
 \sin\theta\sin\phi& \cos\theta\sin\phi  & \cos\phi\\
 \cos\theta& -\sin\theta  & 0 
\end{pmatrix}\, ,
\end{equation}
represents the unitary transformation matrix between the Cartesian and spherical unit vectors $(\hat{x},\hat{y},\hat{z})\to (\hat{r},\hat{\theta},\hat{\phi})$, such that $\nabla = \mathbf{U} \cdot \tilde{\nabla}$ recovers the Cartesian gradient. Explicit analytical expressions of radial integrals and spherical harmonics derivatives are reported in the Supplementary Material.

\subsection{Long-range electronic contribution}

From a smooth choice of $n^\text{ewald}_e$, the long-range part of the Hartree potential can be conveniently written as a plane-waves expansion compatible with the three-dimensional periodicity of the system:
\begin{equation}\begin{split}\label{eq:hartree-LR}
    \phi^\text{LR}_\text{H}(\br) &=\sum_{\bk\ne \boldsymbol{0}}\frac{4\pi}{k^2}\, \tilde{n}^\text{ewald}_e(\bk)\, e^{i\bk\cdot \br} \, .
\end{split}\end{equation}
Note that we excluded the $\bk=0$ term because of the assumed charge neutrality of the overall system, i.e., electrons and nuclei. 
From the expansion of the plane wave in spherical harmonics, the Fourier components of the electron density are found to get the form of
\begin{equation}\begin{split}\label{eq:fourier_components}
    \tilde{n}^\text{ewald}_e(\bk) &= \frac{1}{\Omega} \int_\Omega d \br\, e^{-i\bk \cdot \br} n^\text{ewald}_e(\br)  \\&=
    \frac{4 \pi}{\Omega}\sum_{i=1}^{N_\text{at}} e^{-i\bk\cdot \br_i}\sum_{nlm} c^{nlm}_i (-i)^l Y^{l}_m(\hat{\bk})\,  \hat{I}_{nl}(k)  \, ,
  \end{split}\end{equation}
where $\Omega$ is the cell volume. Here, $\hat{I}_{nl}$ are defined as the following radial integrals,
\begin{equation}
\begin{split}\label{eq:radial_integral_ewald}
    \hat{I}_{nl}(k) &= \int_0^\infty dr\, r^2 j_l(kr)  R^\text{ewald}_{nl}(r)\, ,
    \end{split}
\end{equation}
with $j_l$ a spherical Bessel function of the first kind. Upon plugging Eq.~\eqref{eq:fourier_components} into Eq.~\eqref{eq:hartree-LR}, the final analytical expression for the long-range Hartree potential is

\begin{equation}\begin{split}\label{eq:hartree-final}
    \phi^\text{LR}_\text{H}(\br) &=\frac{32\pi^2}{\Omega} \sum^\text{half}_{\boldsymbol{k}\ne \boldsymbol{0}} \frac{1}{k^2} \sum_{i=1}^{N_\text{at}}\sum_{nlm} c^{nlm}_i  \,  \hat{I}_{nl}(k) \, Y^{l}_m(\hat{\bk}) \\& \times\begin{cases}
      (-1)^{\frac{l}{2}}\cos\left[\bk\cdot (\br-\br_i)\right]\, , \   l\%2=0\\
      (-1)^{\frac{l+3}{2}}\sin\left[\bk\cdot (\br-\br_i)\right]\, , \  l\%2\ne0
      \end{cases}\, .
\end{split}\end{equation}
Note that we exploited the real nature of the electron density to sum $\bk$-vectors only over a semisphere in reciprocal space of radius $k_\text{cut}$. In practice, $k_\text{cut}$ is indirectly defined by the chosen value of $\sigma_\text{Ewald}$, following the numerical convergence criterion implemented in MetalWalls.~\cite{Marin-Lafleche2020} 

Upon Cartesian differentiation of Eq.~\eqref{eq:hartree-final}, the long-range part of the electronic electric field is finally given~by
\begin{equation}\begin{split}\label{eq:electric_field}
    \boldsymbol{E}^\text{LR}_\text{H}(\br) = &\frac{32\pi^2}{\Omega} \sum^\text{half}_{\boldsymbol{k}\ne \boldsymbol{0}} \frac{\bk}{k^2} \sum_{i=1}^{N_\text{at}}\sum_{nlm} c^{nlm}_i\,  \hat{I}_{nl}(k) \, Y^{l}_m(\hat{\bk})  \\& \times \begin{cases}
     (-1)^{\frac{l}{2}}\,  \sin\left[\bk\cdot (\br-\br_i)\right]\, ,\ l\%2=0\\
      (-1)^{\frac{l+1}{2}}\,  \cos\left[\bk\cdot (\br-\br_i)\right]\, ,\ l\%2\ne0
    \end{cases}\, .
\end{split}\end{equation}
An explicit expression of the radial integrals, together with additional implementation details, are reported in the Supplementary Material.

\subsection{Pseudopotential nuclear contribution}

From Eq.~\eqref{eq:pseudo-charge}, computing the nuclear contribution to the electric field is well established,~\cite{marx_hutter_2009} and it reduces to the problem of treating a set of classical Gaussian charges. Following the conventional Ewald method, this amounts to compute a screened short-range potential as
\begin{equation}\label{eq:local-pseudo}
    \phi^\text{SR}_\text{ion}(r) = \frac{Z_\text{eff}}{r} \left[\text{erf}\left(\frac{r}{\sqrt{2}r_\text{loc}}\right)-\text{erf}\left(\frac{r}{\sqrt{2}\sigma_\text{ewald}}\right)\right]\, ,
\end{equation}
and a long-range contribution in reciprocal space, whose Fourier components are given by
\begin{equation}
    \tilde{\phi}^\text{LR}_\text{ion}(k) = \frac{4\pi}{\Omega}\frac{Z_\text{eff}}{k^2} \exp\left(-\frac{1}{2}k^2 \sigma^2_\text{ewald}\right)\ .
\end{equation}
For consistency, we select a value of $\sigma_\text{ewald}$ equivalent to that adopted for the electronic field calculation.

The short-range part of the electric field can be computed by relying on the same transformation matrix between Cartesian and spherical unit vectors already introduced in Eq.~\eqref{eq:unimatrix}. Considering that Eq.~\eqref{eq:local-pseudo} has no angular components, we can write
\begin{equation}\begin{split}
    \boldsymbol{E}^\text{SR}_\text{ion}(\br) = \sum_{i=1}^{N_\text{at}} \sum_{\boldsymbol{u}}\mathbf{U}(\widehat{\br-\br_i-\boldsymbol{u}}) \cdot \left[-\frac{\partial}{\partial r}\phi^\text{SR}_\text{ion}(r),0,0\right]\, .
\end{split}\end{equation}
The long-range part of the nuclear electric field is finally computed as follows,
\begin{equation}\begin{split}
    \boldsymbol{E}^\text{LR}_\text{ion}(\br) &=  2\sum^\text{half}_{\boldsymbol{k}\ne \boldsymbol{0}} \bk\, \tilde{\phi}^\text{LR}_\text{ion}(k) \sum_{i=1}^{N_\text{at}} \sin\left[\bk\cdot(\br-\br_i)\right]\, .
\end{split}\end{equation}
\section{SALTED/MetalWalls interface\label{sec:interface}}

Upon implementing the electric field just derived into the MetalWalls molecular dynamics engine, the SALTED/MetalWalls simulation can be easily run thanks to a Python-based interface between the two programs. In what follows, we report a description of the simulation workflow, including pseudocode of the relevant steps.

\begin{enumerate}
    \item Import SALTED and MetalWalls modules:

    \texttt{import salted, metalwalls}
    
    \item Load pretrained SALTED information:
    
    \texttt{salted\_info = salted.init\_pred.build()}

    \item Load isolated electrode information:

    \texttt{structure = ase.io.read("init.xyz")}\\[0.1cm]
    \texttt{coords\_electrode = structure.get\_positions()[640:]}\\[0.1cm] 
    \texttt{coefs\_0 = np.loadtxt("coefs\_isolated.txt")}

    \item Define MetalWalls variables.
    
    \texttt{my\_system = \\metalwalls.mw\_system.MW\_system\_t()}\\[0.1cm]
    \texttt{my\_algorithms =\\ metalwalls.mw\_algorithms.MW\_algorithms\_t()}\\[0.1cm]
    \texttt{my\_parallel = \\metalwalls.mw\_parallel.MW\_parallel\_t()}

    \item Initialize electrode charge density:
    
    \texttt{coords\_electrolyte = my\_system.xyz\_ions}\\[0.1cm] 
    \texttt{coords = np.vstack((coords\_electrolyte,\\coords\_electrode))} \\[0.1cm]
    \texttt{structure.set\_positions(coords)}\\[0.1cm]
    \texttt{my\_system.coefs = salted.salted\_prediction\\.build(salted\_info,structure)}\\[0.1cm] 
    \texttt{my\_system.coefs += coefs\_0}

    \item Initialize MetalWalls simulation:

    \texttt{metalwalls.mw\_tools.run\_step(my\_system, my\_parallel, my\_algorithms, 0, 0, do\_output, step\_output\_frequency)}

    \item Start the dynamics:

     \texttt{for step in range(my\_system.num\_steps):}

\item Update electrolyte atomic positions:
    
\texttt{metalwalls.mw\_tools.step\_setup(my\_system, my\_algorithms, my\_parallel, step)}\\[0.1cm]
        \texttt{coords\_electrolyte = my\_system.xyz\_ions}\\[0.1cm] 
        \texttt{coords = np.vstack((coords\_electrolyte,\\coords\_electrode))} \\[0.1cm]
    \texttt{structure.set\_positions(coords)}
    
    \item Update electrode charge density:
    
    \texttt{my\_system.coefs = salted.salted\_prediction\\.build(salted\_info,structure)}\\[0.1cm] 
    \texttt{my\_system.coefs += coefs\_0}

    \item Compute atomic forces:
    
     \texttt{metalwalls.mw\_tools.step\_compute\_forces\\(my\_system, my\_parallel, step)}

     \item Continue from point 8.

\end{enumerate}
Importantly, the electrostatic forces acting on the electrolyte atoms $j$ are computed at step 10 as follows,
\begin{equation}\label{eq:forces}
    \boldsymbol{f}_j = q_j\, \boldsymbol{E}(\br_j)\, , 
\end{equation}
where $q_j$ are the electrolyte partial atomic charges, while
\begin{equation}
    \boldsymbol{E} = \boldsymbol{E}^\text{SR}_\text{H} + \boldsymbol{E}^\text{LR}_\text{H} + \boldsymbol{E}^\text{SR}_\text{ion} + \boldsymbol{E}^\text{LR}_\text{ion}\, ,
\end{equation}
is the total electric field generated by the predicted electrode charge distribution, as decomposed from the derivation carried out in Sec.~\ref{sec:efield}.

\section{Results~\label{sec:results}}

\begin{figure}[t]
    \centering
    \includegraphics[width=8.5cm]{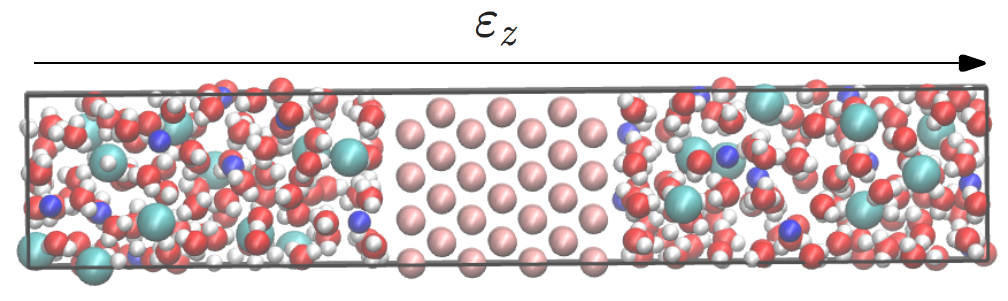}
    \caption{Representation of an ionic capacitor made of a Au(100) gold electrode put in contact with a water/NaCl electrolyte solution under 3D periodic boundary conditions. According to Ref.~\citenum{Dufils2019}, a finite electric field $\varepsilon_z$ is applied along $z$ to mimic a voltage drop of $\Delta V=-\varepsilon_z L_z$ between two parallel electrodes.}
    \label{fig:system}
\end{figure}

To test the simulation workflow previously described, we revisit the example of ionic capacitor introduced in Ref.~\citenum{grisafi2023prm}. In particular, we consider a Au(100) electrode made of 4 unit cell repetitions along the $xy$ plane and 7 atomic layers along~$z$, which is put in contact with an aqueous solution of NaCl 5~M under 3D periodic boundary conditions (Fig.~\ref{fig:system}). The electrolyte is made of 200 water molecules and 20 NaCl ions pairs, which, together with the gold electrode, occupy a box size of $L_{x/y}=11.54$~{\AA} and  $L_z=64.34$~{\AA}. Classical electrolyte interactions are defined using a TIP4P/2005 water model~\cite{Abascal2005} with Na$^+$ and Cl$^-$ Lennard-Jones parameters taken from Ref.~\citenum{Yagasaki2020}; fixed classical atomic charges are given by $q_\text{M}=-1.1128$, $q_\text{H}=+0.5564$ and $q_\text{Na/Cl}=\pm 1.0$. 

Following the discussion reported in Sec.~\ref{sec:method}, we apply a uniform electric field along the direction orthogonal to the Au(100) surface, allowing us to mimic a pair of equivalent gold electrodes that are kept at a constant voltage drop. In particular, we consider two different cell potentials, namely $\Delta V=0$~V and $\Delta V=1$~V, which correspond to applied fields of~$\varepsilon_z=0.0$~V/{\AA} and $\varepsilon_z=0.016$~V/{\AA}, respectively. Reference QM/MM calculations are performed for 2000 uncorrelated frames selected from a classical MetalWalls trajectory obtained running a finite-field simulation at a cell voltage of $1$~V. The basis set used for the gold electron density representation is defined from an uncontracted version of the automatically generated auxiliary functions implemented in CP2K.~\cite{Stoychev2017} In particular, we adopt an auxiliary basis set that includes up to $l_\text{max}=4$ spherical harmonics, accounting for a total of~182 Gaussian-type functions per gold atom. 

\subsection{Validation of SALTED electrode model}

From the QM/MM dataset so generated, SALTED models of the electrode charge density are trained following the discussion of Sec.~\ref{sec:method}-B. In practice, a subset of $M=400$ gold atomic environments is selected to recast the learning problem into a low dimensional space.~\cite{Grisafi2023} Symmetry-adapted kernel functions are computed from LODE descriptors that include both atom-density and potential-like structural features.~\cite{grisafi2023prm,grisafi2021cs} In this work, these are defined using Gaussian widths of $\sigma=0.5$~{\AA} and $\sigma=2$~{\AA}, respectively, together with a radial cutoff of $r_\text{cut}=10$~{\AA}. To validate the model, we select 1600 random configurations for training and retain the remaining 400 for testing. The prediction error is measured consistently with the definition of the SALTED loss function reported in Eq.~\eqref{eq:loss}. In particular, we compute the percentage root mean square error as 
\begin{equation}
   \% \text{RMSE} =  \sqrt{ \frac{\mathcal{L}_\text{test}(\boldsymbol{w})}{ \sigma^2_\text{test}}} \times 100\, ,
\end{equation}
where $\mathcal{L}_\text{test}(\boldsymbol{w})$ follows from Eq.~\eqref{eq:loss}, while
\begin{equation}
    \sigma^2_\text{test} =  \sum_{I=1}^{N_\text{test}}\left[\boldsymbol{c}^\text{DF}_I\right]^T \mathbf{J}_I\,  \boldsymbol{c}^\text{DF}_I\, ,
\end{equation}
represents the variance of the learning target computed over the test set. Learning curves are reported in the Supplementary Material. 

\begin{figure}[t!]
    \centering
    \includegraphics[width=8.5cm]{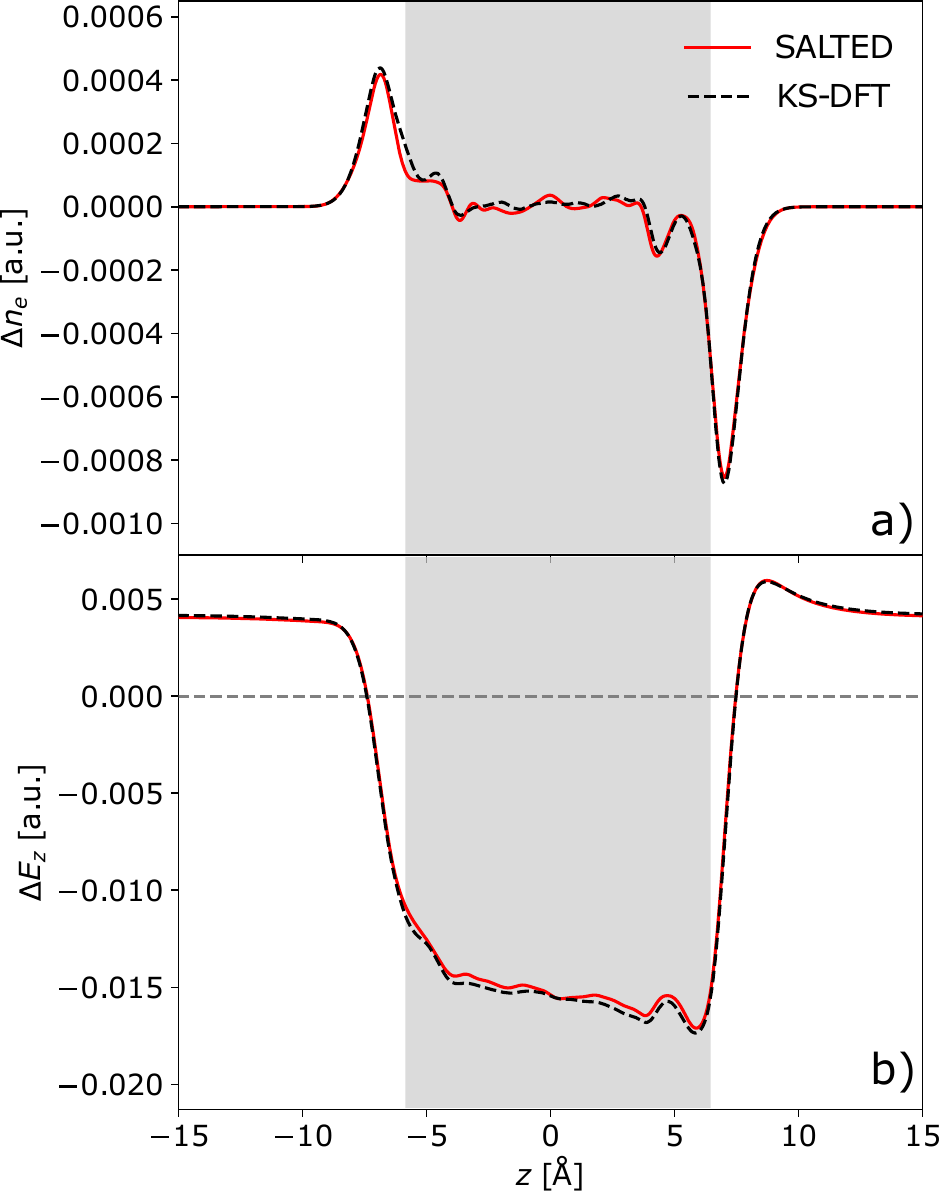}
    \caption{Electrode charge density response (a) and derived electric field (b) induced by the classical charges of a test electrolyte configuration. Red lines: SALTED prediction results. Black dashed lines: Kohn-Sham DFT reference. Shaded gray area indicates the region occupied by the gold electrode.}
    \label{fig:field}
\end{figure}
 
With as few as 200 training structures, the SALTED error saturates at $\sim$3\% RMSE, demonstrating the capability of the model of capturing the non-local fluctuations of the electronic charge induced by  different spatial arrangements of the electrolyte atoms. This result is confirmed by the  excellent agreement between the Kohn-Sham and predicted electron density variation $\Delta n_e$, as depicted in Fig.~\ref{fig:field}-a for a representative test configuration. Despite the slightly less accurate prediction of the charge density profile with respect to what reported in Ref.~\citenum{grisafi2023prm}, we find that the use of a Coulomb metric in the SALTED loss function allows us to obtain a highly accurate electric field. This is illustrated in Fig.~\ref{fig:field}-b, where the longitudinal variation of the predicted electric field with respect to the field of the isolated electrode, $\Delta E_z$, is reported against the quantum-mechanical reference. Notably, we observe that the agreement with DFT is especially good outside the electrode region, being the most relevant for driving the dynamics of the electrolyte.

\begin{figure}[t]
    \centering
    \includegraphics[width=8.5cm]{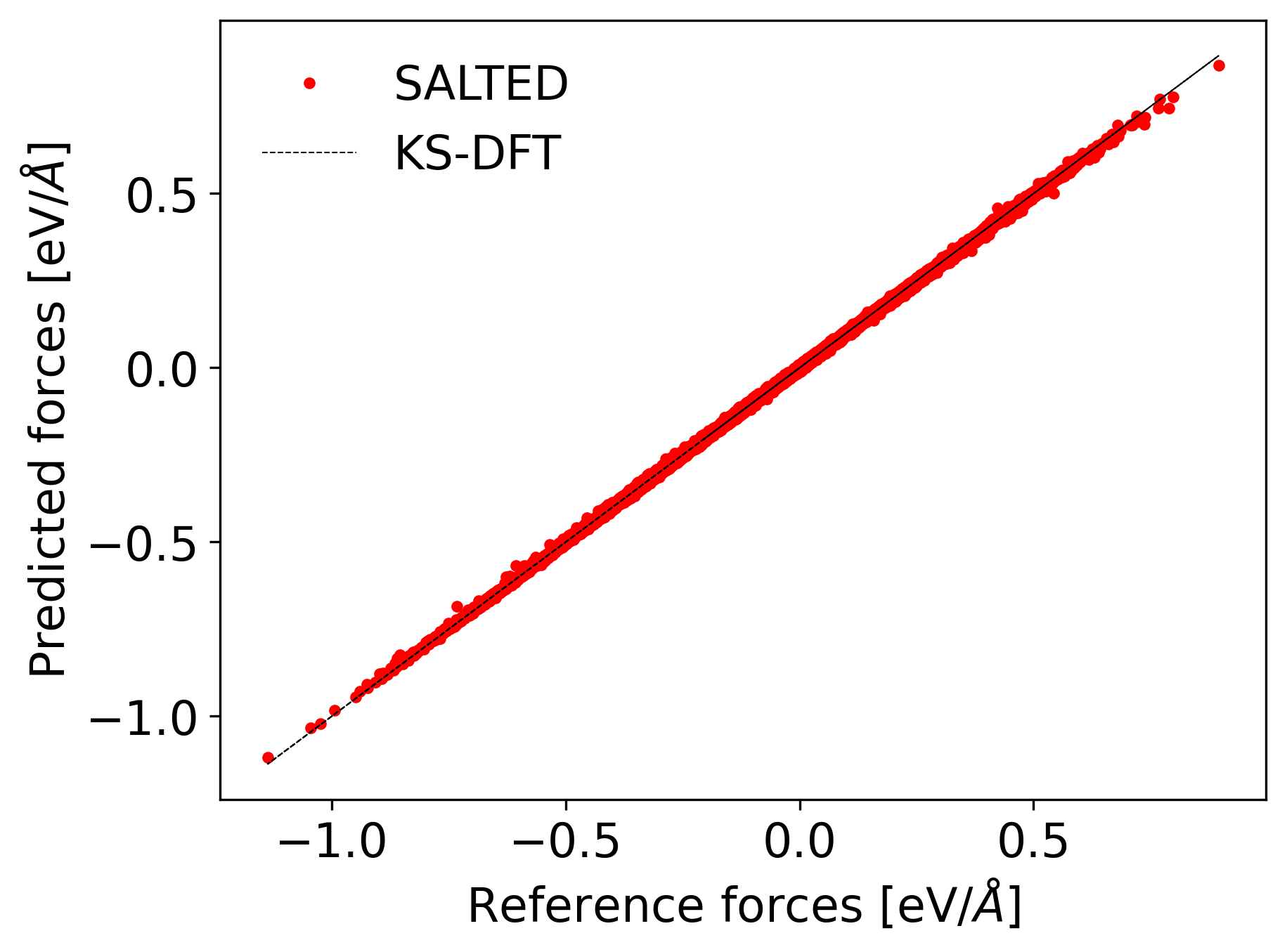}
    \caption{Correlation plot between the predicted and reference electrostatic forces generated by the  electric field of a gold electrode, as computed over 256'000 electrolyte atoms belonging to 400 test configurations. Red dots: SALTED~prediction. Black dashed line: KS-DFT reference. Predicted forces are associated with a $\text{RMSE}=1.0$~meV/{\AA} over a standard deviation of $\sigma_f = 141.9$~meV/{\AA}.}
    \label{fig:forces}
\end{figure}

As an ultimate proof of the model accuracy, we compute the predicted Cartesian components of the electrostatic atomic forces associated with the 400 electrolyte configurations used for testing, as compared against the quantum-mechanical reference. Both reference and predicted forces are computed as in Eq.~\eqref{eq:forces}, starting from the DFT and SALTED electron-density coefficients, respectively. Consistently with the high level of statistical correlation shown in Fig.~\ref{fig:forces}, we obtain a remarkably small root mean square error of $1.0$~meV/{\AA}, which amounts to about 0.7\% of the standard deviation of the DFT forces in the test~set. By and large, these results highlight the intrinsic effectiveness of a machine learning model that has the electron density as a prediction target to compute the electrostatic properties of the system.

\subsection{Simulation of ionic capacitor}
We now report the results of the data-driven molecular dynamics, which is run following the simulation workflow described in Sec.~\ref{sec:interface}. Finite-field simulations associated with applied cell potentials of $\Delta V =0$~V and $-$1~V are performed at room temperature in timesteps of 2~fs using a Nos\'e-Hoover thermostat.~\cite{nose84jcp} Predictions of the electrostatic atomic forces are obtained from a SALTED model trained on $N=2000$ configurations, as validated from the discussion previously carried out. By~parallelizing the calculation over an Intel i9-12900 CPU, we find that each step of the dynamics takes about 0.5~s, corresponding to a speedup $>10^3$ with respect to first-principles QM/MM calculations.
This substantial increase in computational efficiency results to be decisive for achieving timescales of the order of nanoseconds, thus making it possible to access the slow dynamics of the electrolyte while preserving a highly accurate representation of electrode charge density. For both applied voltages, in particular, we are able to reach long simulation times of $\approx 3$~ns. 

It is instructive to compare our results  against those of a classical MetalWalls simulation. In this case, the electronic polarization of the gold electrode is approximated through a charge-equilibration scheme that makes use of fluctuating Gaussian charges.~\cite{Marin-Lafleche2020} Specifically, we represent the partial charges on the gold atoms through a Gaussian width of $\sigma_\text{MW}=1.06$~{\AA}, and simulate classical trajectories for a total time that is, once again, of about 3~ns. We note that similar simulations are only 50 times faster than those performed through our method. This is remarkable if considering that, in addition to the SALTED predictions, quantum electrostatic forces are computed from the charge density generated by 182 spherical harmonics functions per gold atom, rather than from a single isotropic Gaussian function.

\begin{figure}[t]
    \centering
    \includegraphics[width=8.7cm]{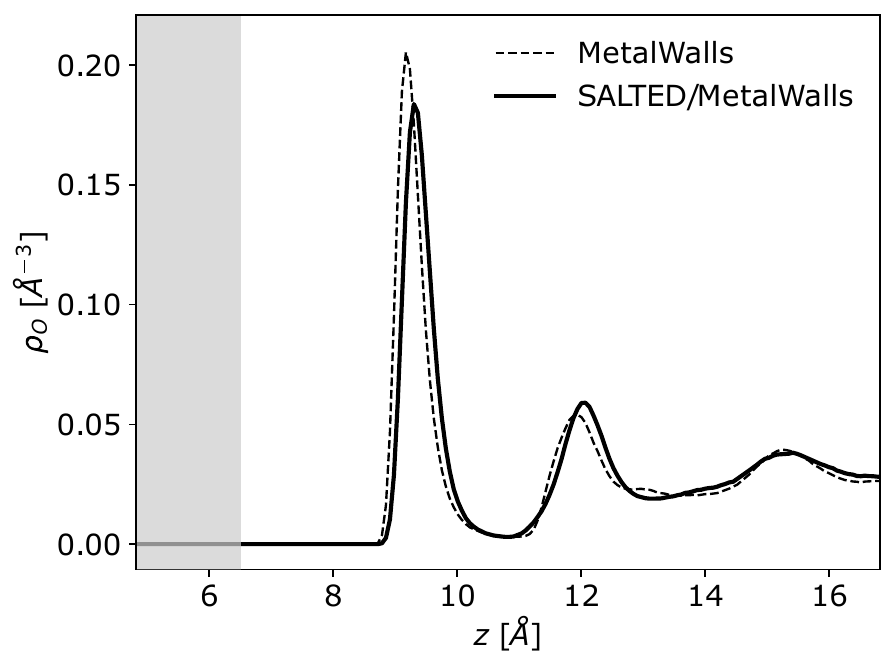}
    \caption{Longitudinal density profiles of oxygen atoms on the right hand side of the ionic capacitor studied in this work, subject to a null cell potential of $\Delta V=0$~V. Dashed  line: classical MetalWalls simulation. Full line: SALTED/MetalWalls simulation.  Shaded gray area indicates the gold electrode.}
    \label{fig:oxygen_right_0V}
\end{figure}

Regardless from the applied potential, we find that the thermal distribution of water molecules obtained through our method displays a systematic shift with respect to the result of a classical simulation. In particular, we observe a distancing of about 0.5~{\AA} of the first adsorption peak, which comes along with overall smoother density profiles. This is shown in Fig~\ref{fig:oxygen_right_0V} in the case of a vanishing applied voltage, indicating that a  classical electrode model tends to give a stiffer potential of mean force with respect to what expected from a first-principles approach.

\begin{figure}[t]
    \centering
    \includegraphics[width=8.7cm]{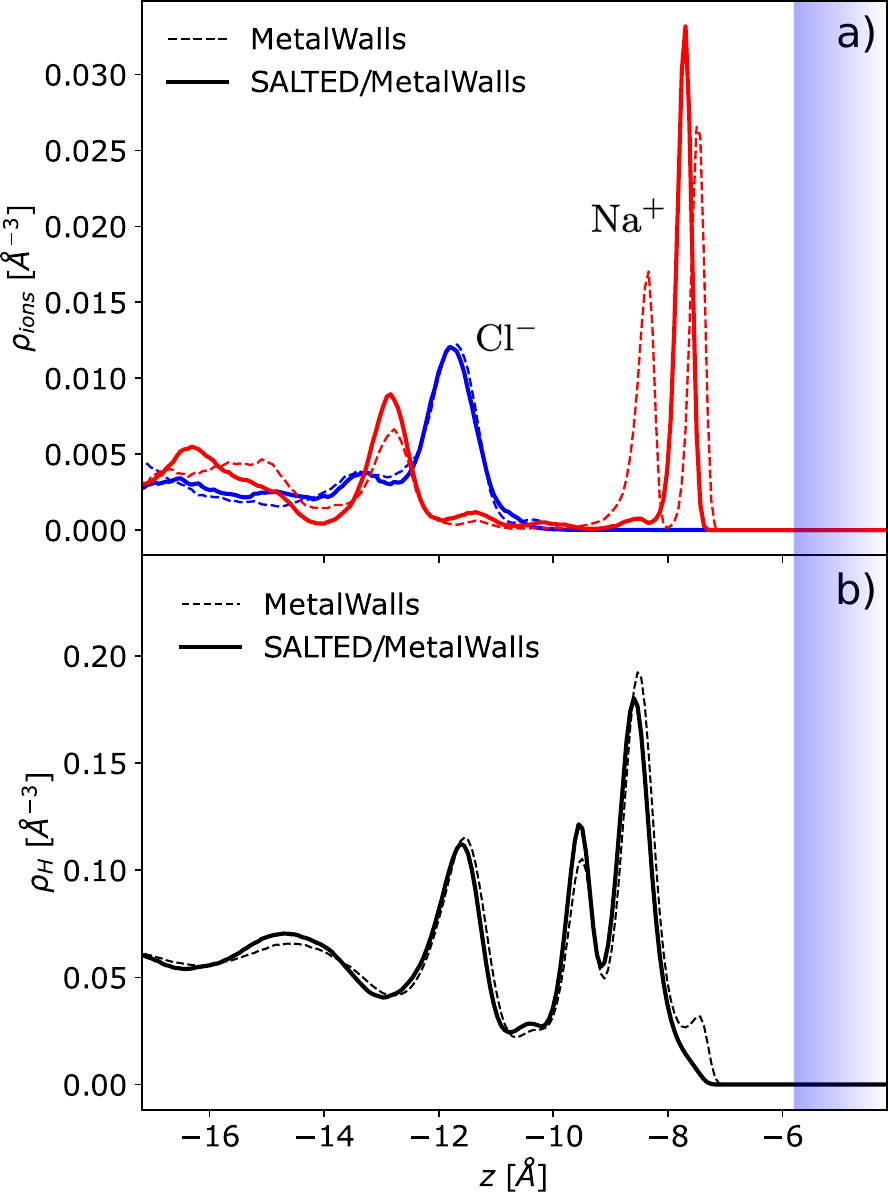}
    \caption{Longitudinal density profiles of sodium chloride (a), and hydrogen atoms (b) at the negatively charged electrode surface of the ionic capacitor studied in this work, subject to an applied cell potential of $\Delta V=1$~V. Dashed lines: classical MetalWalls simulation. Full lines: SALTED/MetalWalls simulation. Red and blue color codes refer to Na$^+$ and Cl$^-$ ions, respectively. Shaded blue area indicates the  gold electrode.}
    \label{fig:ions_left_1V}
\end{figure}

More striking differences between the classical and data-driven simulations emerge when looking at the results obtained under an applied potential $\Delta V=1$~V. As an example, we report in Fig.~\ref{fig:ions_left_1V}-a the ionic density profiles of Na$^+$ and Cl$^-$ associated with the screening of the negatively charged metal surface. While the distribution of chlorine shows an overall agreement between the two methods, a qualitative difference is found in the adsorption of sodium cations. In particular, a double pick is produced as a result of the classical simulation, which does not appear when using SALTED to predict the quantum-mechanical response of the electrode.  This artifact of the classical model can be associated with the rigidity by which the distribution of the metal charge is represented, which mostly affects the accuracy of the electric field in the proximity of the gold electrode. In fact, it has already been shown that, depending on the chosen value of  $\sigma_\text{MW}$, MetalWalls simulations can yield substantial variations in the distribution of Na$^+$ at the Au(100) surface.~\cite{Serva2021}

In addition to the ionic profiles, a clear-cut discrepancy is also observed in the distribution of hydrogen atoms; this is illustrated in Fig.~\ref{fig:ions_left_1V}-b. Interestingly, we find that a prepeak is produced by the classical simulation at about 2~{\AA} from the gold slab. This prepeak is associated with an occasional reorientation of the water molecules that expose the hydrogen atoms towards the electrode, a phenomenon that is however not predicted by our method. This is confirmed by an analysis of the orientation of the water dipoles with respect to the normal to the electrode surface, as reported in the Supplementary Material.

\begin{figure}[t]
    \centering
    \includegraphics[width=8.7cm]{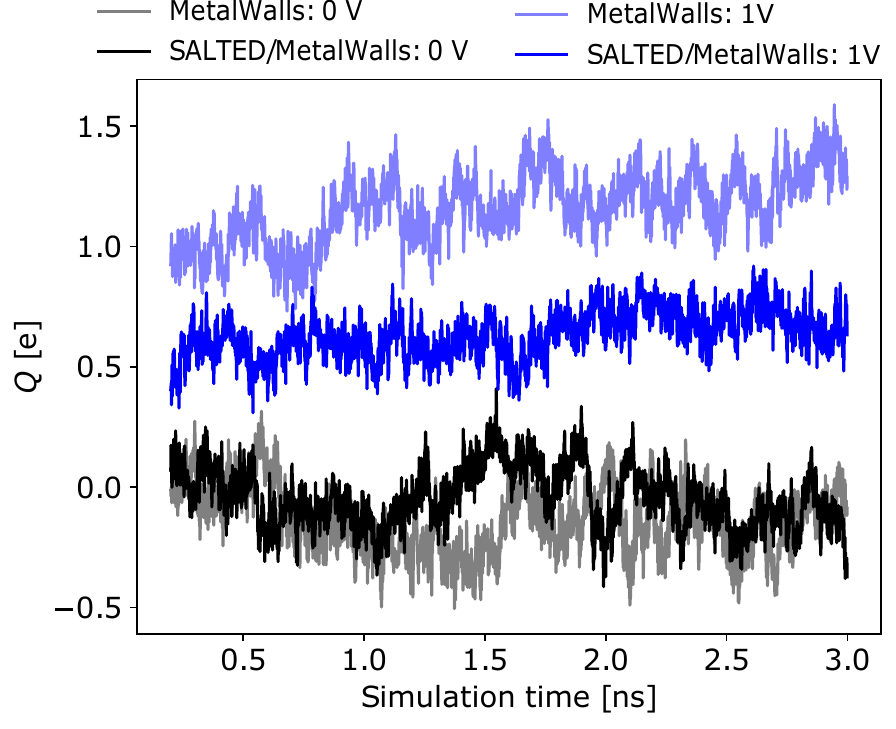}
    \caption{Evolution of the electrode surface charge in the ionic capacitor studied in this work. Black and blue lines refer to SALTED/MetalWalls simulations at 0 and $1$~V, respectively. Gray and light blue lines refer to classical MetalWalls simulations at 0 and $1$~V, respectively.}
    \label{fig:charge_fluctuations}
\end{figure}

A further benefit of driving the dynamics of the electrochemical interface through electron-density predictions is that of giving access to the total charge $\pm Q$ accumulated at the electrode surfaces. 
In a finite-field simulation setup, $Q$ can be computed by integrating  $\Delta n_e$ from the middle of the metal slab to the classical electrolyte region.~\cite{grisafi2023prm} In Fig.~\ref{fig:charge_fluctuations}, we report the time evolution of $Q$, computed every picosecond, for both null and finite cell potentials. When considering the results at $\Delta V=0$~V, we observe a fluctuation of the electrode charge around 0 for both classical and SALTED-based simulations. This is to be expected from the symmetry of the isolated metallic slab, implying that no net polarization can be found at thermal equilibrium without an externally applied electric bias. Conversely, simulations run at $\Delta V=1$~V lead to a finite average charge $\langle Q\rangle$. In particular, we find that the computed classical value of $\langle Q\rangle$ is about twice as large than what predicted through our method, i.e., 1.16$e$ versus 0.63$e$, respectively. This difference in the system's electric capacitance is due to a substantial overestimation of the electronic polarization of the metallic slab obtained through the classical model, a result that was already hinted in Ref.~\citenum{grisafi2023prm} when comparing predictions of $\Delta n_e$ performed on a test classical trajectory.
Although a better agreement might be obtained by tuning the value of the classical Gaussian width $\sigma_\text{MW}$, the observed discrepancy suggests that disposing of a physically sound electrode model is preferable to obtain reliable results.

In addition to $\langle Q\rangle$, having access to the thermal fluctuations of $Q$ allows us to provide an estimate of the differential capacitance associated with the formation of the electrical double layer:~\cite{Scalfi2020}
\begin{equation}
    C^\text{EDL}_\text{diff} = \beta \left<(Q-\langle Q\rangle)^2\right>\, .
\end{equation}
At $\Delta V=0$~V, a similar differential capacitance is obtained for both type of simulations, resulting in a value of $C^\text{EDL}_\text{diff}=8.9 \mu$F/cm$^2$ for the classical model, and a slightly reduced value of  $C^\text{EDL}_\text{diff}=7.4 \mu$F/cm$^2$ for the data-driven model. This discrepancy is magnified when looking at the results at $\Delta V=1$~V. In line with what already computed in Ref.~\citenum{grisafi2023prm}, the classical simulation is associated with a value of $C^\text{EDL}_\text{diff}=10.1 \mu$F/cm$^2$. Conversely, our SALTED/MetalWalls simulation comes along with much more attenuated charge fluctuations, yielding a differential capacitance that is about twice as small, i.e., $C^\text{EDL}_\text{diff}=4.6 \mu$F/cm$^2$. While simulations longer than 3~ns would be required to provide a tighter convergence of $C^\text{EDL}_\text{diff}$, these results remark, once again, the need of introducing an accurate representation of the electrode in the description of the electrochemical interface.

\section{Conclusions}

The presented method represents a relevant example of how to incorporate a machine-learning approach that addresses the prediction of electronic-structure properties into the atomistic study of complex materials. By providing a formally rigorous integration between the SALTED and MetalWalls programs, we have shown how to extend the reach of applications of QM/MM approaches that aim at preserving a quantum description of metal electrodes in the simulation of electrochemical interfaces. Most notably, the possibility of driving the dynamics of the electrolyte over nanoseconds timescales represents a great leap forward towards obtaining a thermally relaxed, yet accurate, description of the electrical double layer. 

By taking an ionic capacitor as an example, we have found that disposing of a quantum treatment of the electrode charge density represents a critical aspect for predicting the structural properties of the electrolyte at the interface. In particular, our results remark how a classical electrode model cannot reproduce the complex nature of the system's electrostatics close to the metal surface. This is not surprising considering that strong electric fields are known to be encountered whenever an electrolyte atom penetrates the electronic cloud of the metal electrode,~\cite{Schmickler1996} an effect that is especially difficult to include through the use of classical atomic charges.  In this regard, an interesting question will be that of understanding how our method compares against data-driven charge equilibration schemes based on partial charges, where a ML representation of atomic electronegativies,~\cite{Ko2021} electric dipoles,~\cite{Staacke2022} and/or  polarizabilities,~\cite{Shao2022} is adopted to predict the electrostatics of the system. We mention, in particular, a recent ML interface with MetalWalls that can be used to refine the atomic Gaussian charges of the electrode by relying on a linear density-response formalism.~\cite{Dufils2023}

In perspective, a notable extension of our method could involve promoting redox active molecular species to the QM region, thus enabling the simulation of electrochemical reactions in the presence of an explicit ionic solution. In this context, an important  aspect will be that of integrating our approach with already existing local ML potentials. This possibility would be especially attractive both for studying the charging mechanisms in pseudocapacitive materials,~\cite{Choi2020} as well as for rationalizing the role of the electrolyte in electrocatalytic processes.~\cite{ringe2019} Furthermore, disposing of a three-dimensional map of the electronic charge density will constitute an added value of our method in view of predicting the regioselectivity of the catalyst to electron-transfer phenomena.

\section*{Supplementary Material}
See the Supplementary Material for a complete derivation of the electronic electric field, together with complementary results.

\section*{Author Declarations}
The authors have no conflicts to disclose.

\section*{Source code}
Printing of density coefficients and 2-center auxiliary integrals is made available under the official trunk version of CP2K.~\footnote{\url{https://github.com/cp2k/cp2k}} The latest SALTED implementation used to produce the results of this work can be accessed from GitHub.~\footnote{\url{https://github.com/andreagrisafi/SALTED}} A version of MetalWalls suitable to be interfaced with SALTED can be downloaded from GitLab.~\footnote{\url{https://gitlab.com/andreagrisafi/metalwalls/-/tree/salted_interface?ref_type=heads}}

\section*{Data availability}
The data that support the findings of this study are openly available from Zenodo.~\footnote{\url{https://zenodo.org/doi/10.5281/zenodo.11175494}} 

\begin{acknowledgments}
We thank Alessandra Serva, Laura Scalfi and Augustin Bussy for useful discussion. This work was supported by the French National Research Agency under the France 2030 program (Grant ANR-22-PEBA-0002). 
\end{acknowledgments}

\end{document}